\documentclass[conference]{IEEEtran}
\IEEEoverridecommandlockouts
\usepackage{cite}
\usepackage{amsmath,amssymb,amsfonts}
\usepackage{algorithmic}
\usepackage{graphicx}
\usepackage{textcomp}
\usepackage{xcolor}

\usepackage[hyphens]{url}  
\usepackage{hyperref}
\usepackage{booktabs}
\usepackage{multirow}

\def\BibTeX{{\rm B\kern-.05em{\sc i\kern-.025em b}\kern-.08em
    T\kern-.1667em\lower.7ex\hbox{E}\kern-.125emX}}
\begin{document}

\title{dLLM-ASR: A Faster Diffusion LLM-based Framework for Speech Recognition}


\author{
\IEEEauthorblockN{
Wenjie Tian\IEEEauthorrefmark{2}, Bingshen Mu\IEEEauthorrefmark{2}, Guobin Ma\IEEEauthorrefmark{2}, Xuelong Geng\IEEEauthorrefmark{2},  Zhixian Zhao\IEEEauthorrefmark{2},  Lei Xie\IEEEauthorrefmark{1}\IEEEauthorrefmark{2}
}

\IEEEauthorblockA{
\IEEEauthorrefmark{2}Northwestern Polytechnical University, Xi’an, China \\
twj@mail.nwpu.edu.cn, lxie@nwpu.edu.cn
}

\thanks{\IEEEauthorrefmark{1}Corresponding author.}
}
\maketitle

\begin{abstract}
Automatic speech recognition (ASR) systems based on large language models (LLMs) achieve superior performance by leveraging pretrained LLMs as decoders, but their token-by-token generation mechanism leads to inference latency that grows linearly with sequence length.
Meanwhile, discrete diffusion large language models (dLLMs) offer a promising alternative, enabling high-quality parallel sequence generation with pretrained decoders.
However, directly applying native text-oriented dLLMs to ASR leads to a fundamental mismatch between open-ended text generation and the acoustically conditioned transcription paradigm required by ASR. 
As a result, it introduces unnecessary difficulty and computational redundancy, such as denoising from pure noise, inflexible generation lengths, and fixed denoising steps.
We propose dLLM-ASR, an efficient dLLM-based ASR framework that formulates dLLM's decoding as a prior-guided and adaptive denoising process. It leverages an ASR prior to initialize the denoising process and provide an anchor for sequence length. Building upon this prior, length-adaptive pruning dynamically removes redundant tokens, while confidence-based denoising allows converged tokens to exit the denoising loop early, enabling token-level adaptive computation.
Experiments demonstrate that dLLM-ASR achieves recognition accuracy comparable to autoregressive LLM-based ASR systems and delivers a 4.44$\times$ inference speedup, establishing a practical and efficient paradigm for ASR.

\end{abstract}


\section{Introduction}
\label{sec:intro}

In recent years, large language models (LLMs)~\cite{llama3, qwen3} have demonstrated unprecedented capabilities in language understanding and generation. This success has catalyzed a paradigm shift in automatic speech recognition (ASR), leading to a unified architecture comprising a speech encoder and an LLM decoder.
LLM-based ASR (LLM-ASR) systems~\cite{whisper, qwenaudio2, whisperasr1, miniomni2, qwen25omni, freeomni}  benefit significantly from scaling laws, achieving superior performance on massive datasets by leveraging the strong linguistic capabilities of the pretrained LLM backbone. Despite their high recognition accuracy, LLM-ASR models predominantly rely on the autoregressive (AR) generation mechanism. This token-by-token decoding process results in inference latency that grows linearly with sequence length $\mathcal{O}(N)$, posing a critical bottleneck for real-time applications.

Conversely, non-autoregressive (NAR) ASR models~\cite{ctcasr1, maskctc, ctcdecoder, trasnducer1, trasnducer2, paraformer} offer faster, parallel inference $\mathcal{O}(1)$ but often suffer from performance degradation. 
One of the important factors is that end-to-end NAR approaches rarely incorporate large-scale pretrained LLMs, due to the inherent mismatch between the NAR modeling and the AR pretraining paradigm.
Consequently, they lack the robust semantic reasoning and rich world knowledge from foundational models, limiting their performance on syntactically complex sentences, rare entities, or context-dependent homophones.

Recently, discrete diffusion LLMs (dLLMs)~\cite{dream, llada, lladav1, lladav2, lladav3, dllm_first, lladav5, lladav6} have introduced a new perspective by modeling the token generation process as parallel iterative denoising in a discrete space. 
dLLMs abandon the left-to-right dependency constraint but retain the powerful semantic capabilities learned by massive pretraining, enabling the parallel generation of the entire sequence within $\mathcal{O}(K)$ steps. 
Specifically, they learn to reverse a process that incrementally corrupts data with noise, achieving a coarse-to-fine generative approach.
This naturally leads to a research question: \textit{\textbf{Can we introduce dLLMs into ASR to achieve both AR-level accuracy and NAR-level inference speed within a unified framework?}}

Although dLLMs are theoretically efficient, native implementations often make their actual inference speed slower than AR models in practice due to redundant computation.
While recent works~\cite{d1, fastdllm, lladav2, dllmvar} have attempted to accelerate text-based dLLMs, directly transferring these techniques to ASR does not always work. This is because of the mismatch that dLLMs are designed for open-ended text generation where content is unconstrained, whereas speech recognition is inherently a mapping task where the output is tightly constrained by the acoustic input. 
As a result, it introduces unnecessary difficulty and computational redundancy when applying dLLMs to ASR.
For example, standard denoising in dLLM typically starts from a fully masked sequence, ignoring the rich and readily available acoustic priors in the input speech.
Further, dLLMs directly applied in ASR suffer from inflexibility as they require a preset generation length. Over-allocating the length leads to redundant padding while under-allocation results in incomplete outputs.
In addition, existing dLLMs~\cite{llada, mmada, speechdllm1, speechdllm2} adopt a uniform compute budget strategy, applying the same number of denoising steps to all tokens. Nevertheless, the difficulty of predicting different tokens varies especially in ASR. Easy words or tokens with clear acoustic features often converge early during denoising, whereas the others require more steps. The one-size-fits-all schedule results in severe computational redundancy.

To address these challenges, we propose dLLM-ASR, a concrete diffusion LLM-based framework that transforms the recognition process to a prior-guided and adaptive denoising process.
Structurally, dLLM-ASR comprises three key components: a speech encoder for acoustic perception, a lightweight adapter for modality alignment, and a pre-trained dLLM serving as the decoder. 
To adapt the dLLM to the speech domain while fully leveraging its inherent world knowledge, we employ a two-stage training strategy to keep the pretrained capability.
Instead of denoising from pure noise, dLLM-ASR utilizes a lightweight ASR prior to guide the denoising process. This serves two critical roles: (1) it acts as a semantic anchor, allowing the model to focus on refining ambiguous regions rather than guessing from scratch; and (2) it provides an explicit reference for sequence length.
Under this prior-guided paradigm, we implement token-level adaptive diffusion that length-adaptive pruning dynamically removes redundant padding, while confidence-based denoising allows converged tokens to exit the denoising loop early.

In summary, our main contributions are summarized as follows:
\begin{itemize}
    \item We propose dLLM-ASR, a faster and more efficient dLLM-based framework tailored for speech recognition.
    \item We investigate the mismatch between text-oriented dLLMs and acoustically conditioned ASR, making discrete diffusion decoding more practical for ASR by introducing a prior-guided and adaptive denoising process.
    \item Experiments demonstrate that dLLM-ASR achieves speech recognition performance comparable to LLM-ASR models while delivering a $4.44\times$ speedup, establishing a practical and efficient paradigm for ASR.
\end{itemize}

\section{Related Work}

\subsection{Automatic Speech Recognition Models}
The development of ASR models is largely defined by the trade-off between inference efficiency and linguistic capability.
NAR-based models aim for fast inference speed through parallel generation. Early works rely on connectionist temporal classification (CTC)~\cite{ctcasr1, maskctc, ctcdecoder}, predicting tokens independently based on frame-level alignments. Subsequent CIF-based models~\cite{paraformer} introduce continuous integrate-and-fire mechanisms to explicitly predict generation length. 
However, without the support of large-scale foundation models, they suffer from restricted linguistic modeling capabilities.
Conversely, LLM-based ASR models~\cite{llmasr1,whisperasr1,whisper,llmasr2, llmasr3} employ a speech encoder to perceive speech information and a modality alignment projector to map speech representations into the LLM's semantic space. They excel at recognizing mixed-language speech, handling long-tail vocabulary, and other paralinguistic tasks.
Despite their linguistic superiority, they face significant hurdles regarding the inefficient token-by-token AR decoding and hallucinations. 
Therefore, there remains a critical need for a framework that unites the parallel efficiency of NAR models with the linguistic power of LLMs.

\subsection{Discrete Diffusion LLMs}
Recently, dLLMs have introduced diffusion mechanisms into the discrete text space. SEDD~\cite{sedd} establish the theoretical foundation, while LLaDA~\cite{llada} and Dream~\cite{dream} model generation as iterative denoising from a fully masked sequence. Although promising, they are often slower than AR LLMs. This is mainly due to high computational overhead, particularly overlong preset generation lengths and numerous denoising steps.
D1~\cite{d1} and LLaDA-1.5~\cite{lladav2} explored reducing inference steps via distillation or reinforcement learning. dLLM-VAR~\cite{dllmvar} achieves variable generation lengths through chunk-wise autoregression. 

We note that several recent works~\cite{speechdllm1, speechdllm2, flowmatchingasr} have explored the application of dLLMs to speech. 
However, these studies primarily investigate the feasibility of dLLM in ASR and overlook the domain-specific challenges.
dLLM-ASR is distinct in that it reformulates ASR decoding as a prior-guided and adaptive denoising process, not the vanilla open-ended generation.
This perspective enables a more favorable trade-off between recognition accuracy and inference latency.

\begin{figure*}[t!]
  \centering
  \includegraphics[width=1\linewidth]{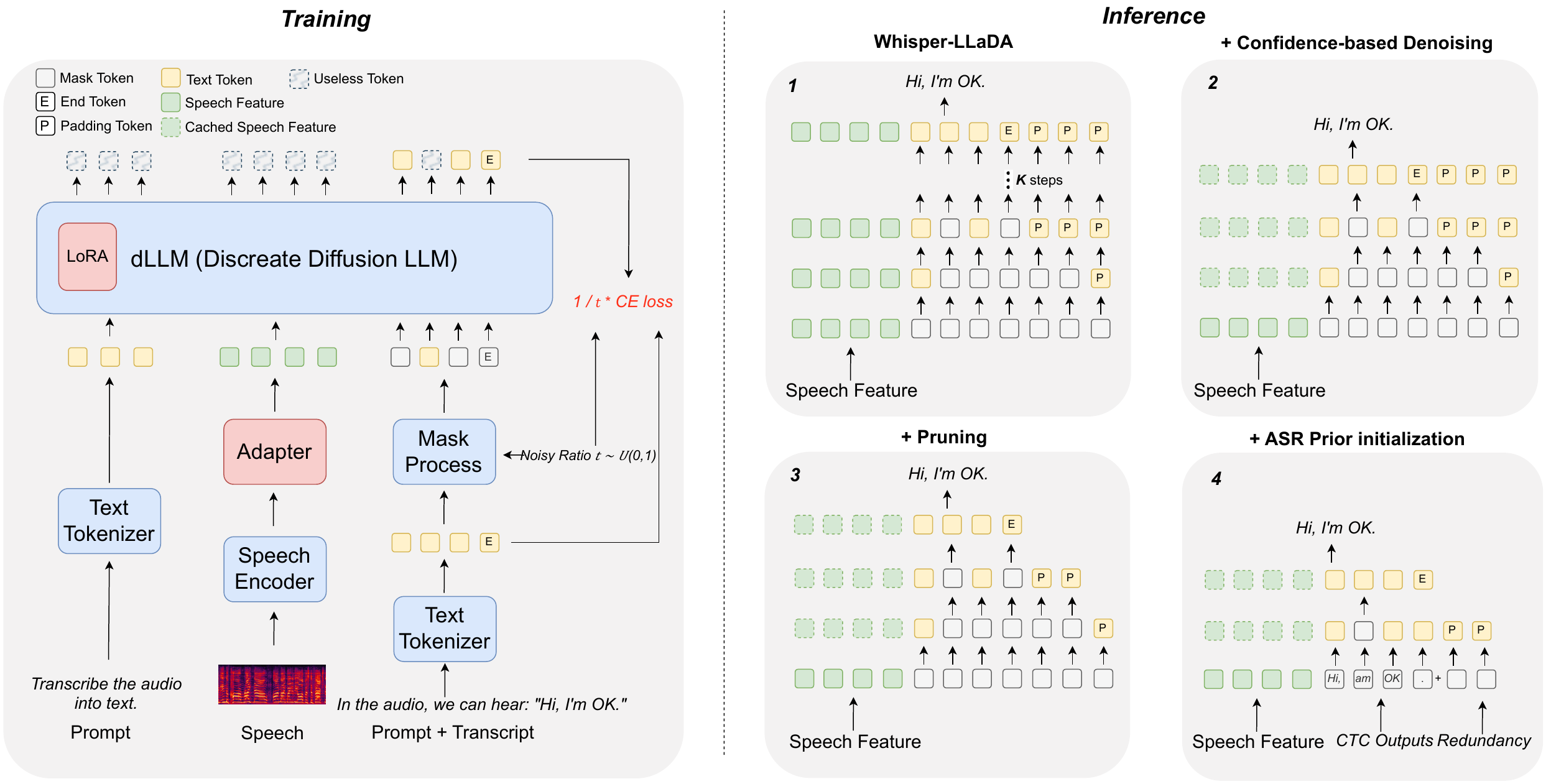}  
  \caption{Overview of the proposed dLLM-ASR.
  On the left side of the figure is the model's architecture and training process. The input consists of the prompt, input speech, and answer, where the answer comprises the prompt and the ground-truth speech transcript. Red components indicate trainable modules, while blue ones represent frozen modules. The right side displays the transformation of the inference process. Subfigure 1 shows the baseline Whisper-LLaDA, which directly combines the speech encoder with LLaDA for ASR. Subfigures 2–4 visualize how our proposed method is built by incrementally applying confidence-based denoising, pruning, and ASR Prior initialization.}
  \label{fig_speechdllm_overview}
\end{figure*}

\section{dLLM-ASR Framework}

\subsection{Overall Architecture}
As illustrated on the left of Fig.~\ref{fig_speechdllm_overview}, our model architecture is composed of three primary components: a speech encoder for robust acoustic perception, an adapter for modality alignment, and a dLLM that serves as the generative backbone.

\subsection{Speech Encoder}
To ensure comprehensive speech understanding, we utilize a powerful pretrained speech encoder. The encoder operates with an output frame rate of 25 Hz.
Following the design in~\cite{stepaudio, whisperasr1}, the encoder remains frozen throughout the training phases.
This strategy preserves the rich semantic and paralinguistic representations learned from massive pre-training data while maintaining training efficiency.

\subsection{Adapter}
To bridge the modality gap, we employ a lightweight adapter that consists of 1D convolutional and linear layers. This adapter maps the acoustic features to the dLLM’s input dimension and reduces the frame rate, yielding a more compact sequence for efficient modeling.
Let $W$ denote the raw input speech, and the final acoustic conditioning features $A$ are obtained as:
\begin{equation}
A = \text{Projector}\big(\text{SpeechEncoder}(W)\big).
\end{equation}

\subsection{dLLM Decoder}
As shown on the left of Fig.~\ref{fig_speechdllm_overview}, the dLLM decoder receives three primary inputs: the text prompt, the speech to be transcribed, and the ground-truth transcript. 
We employ LLaDA~\cite{llada}, a pretrained text dLLM model, as the backbone of our decoder. In text dialogue systems, dLLM models the generation process as conditional denoising, in which the input prompt serves as the context for generating the response. Similarly, in dLLM-ASR, the input speech features serve as the prompt to generate the ASR transcription. Specifically, the speech features $A$ are concatenated with the text embeddings to produce a speech-conditioned multimodal input.

During training, unlike AR models that predict the next token sequentially, dLLM-ASR learns to recover missing information from a partially noised sequence. Let $x_0$ be the ground-truth ASR transcript. We define the forward masking process that transforms $x_0$ into a noisy sequence $x_t$. In this process, tokens in $x_0$ are independently replaced by a special mask token $[\mathrm{M}]$ with a probability $t \in (0, 1]$. The training objective is to predict the original unmasked tokens based on the noisy sequence $x_t$ and the speech condition $A$. The model acts as a denoiser, maximizing the probability of the true tokens at the masked positions. The loss $\mathcal{L}$ is defined as:
\begin{equation}
    \mathcal{L} = -\mathbb{E}_{t, x_0, x_t} \left[ \frac{1}{t} \sum_{i=1}^{L} \mathbb{I}(x_t^i = [\mathrm{M}]) \log p_{\theta}(x_0^i | x_t, A) \right],
\end{equation}
\noindent
where $L$ is the sequence length, and the term $\mathbb{I}(x_t^i = [\mathrm{M}])$ ensures that the loss is calculated only on the masked tokens. By conditioning on $A$, the dLLM learns to align acoustic cues with textual semantics to reconstruct the transcript. To mitigate the bias caused by the varying number of masked tokens across different time steps, we introduce a scaling factor $\frac{1}{t}$ . This term normalizes the objective function, ensuring that when fewer tokens are masked, the loss is assigned a larger weight to balance the optimization process.
What's more, to improve the robustness of the model, we set the noise level to its maximum, i.e., $t=1$, with probability $\alpha$.

\subsection{Training Strategies}
\label{training_strategy}

To effectively align the modalities and adapt the model, following\footnote{\url{https://github.com/k2-fsa/icefall/tree/master/egs/speech_llm/ASR_LLM}}, we employ a two-stage training strategy.
In the first stage, we train only the adapter while keeping the speech encoder and dLLM decoder frozen. 
In the second stage, we introduce Low-Rank Adaptation (LoRA)~\cite{lora} modules into the dLLM backbone. Both the adapter and the LoRA parameters are optimized jointly. 
Interestingly, in our experiments, we observe that the data format plays a pivotal role in generation quality. We wrap the speech representations and transcripts into a chat-style format mimicking text LLMs by adding natural prompts. This strategy appears to activate the pretrained model's instruction-following capabilities by bridging the cross-modal gap.

\subsection{Prior-Guided Adaptive Denoising}
\label{sec:adaptive_decoding}
We propose to reformulate the vanilla inference process as a prior-guided adaptive denoising process. This paradigm shifts the focus from blind generation to efficient refinement based on token-level difficulty.

\subsubsection{Confidence-based Denoising}
Instead of enforcing a fixed number of denoising steps for all tokens, we introduce the mechanism to monitor the confidence score at each step. Once a token's confidence exceeds a pre-defined threshold~$\tau$, we skip its subsequent denoising steps. This early-exit strategy significantly reduces computational overhead by avoiding redundant processing on ``easy" tokens, ensuring efficient inference without compromising recognition quality.
To mitigate premature commitment to incorrect tokens, the mechanism is applied conservatively with a high threshold. 

\subsubsection{Length-Adaptive Pruning}
We empirically observe that the generation length is a critical factor affecting both computational efficiency and token convergence speed.
And in ASR, simple tokens, particularly padding tokens, exhibit low entropy and significantly faster convergence than semantic content, making them particularly suitable for early pruning.
Leveraging this, dLLM-ASR can explicitly identify the majority of these ``easy" tokens at the very first denoising iteration.
By synergizing with confidence-based denoising, we prune these high-confidence padding tokens at the onset, dynamically establishing a tighter upper bound for the sequence length. 
This strategy is applied iteratively, eliminating computational waste and accelerating the overall inference process.

\subsubsection{Efficient Speech Cache}
Inspired by~\cite{fastdllm}, we save the KV cache corresponding to the speech features and reuse it for subsequent denoising steps. Although KV cache is theoretically imprecise in an NAR context, we observe almost no performance degradation when the cache is strictly limited to the speech features. 

\subsubsection{Prior Initialization}
We leverage a lightweight CTC branch trained on the frozen speech encoder output $A$ to generate an initial ASR prior.
Notably, the additionally learnable branch introduces negligible computational overhead, as it's composed of only a downsampling convolution and a classification head.
Serving as the starting state of the denoising process, this prior provides richer contextual information at the first denoising step. 
Consequently, it yields a higher number of tokens surpassing the confidence threshold $\tau$ compared to starting with a fully masked sequence. 
In addition, the prior naturally provides a reference for generation length.

Overall, the evolution of our inference process is illustrated from subfigure 1 to subfigure 4 in the right panel of Fig.~\ref{fig_speechdllm_overview}. During inference, we initially utilize the ASR prior to initialize both the starting denoising state and the generation length.
During the first denoising iteration, we fix the candidate tokens that have a confidence greater than $\tau$, extract the speech KV cache, and detect the end-of-sequence padding.
For all subsequent steps, we reuse the cached speech KV pairs and only update the undecided positions, removing pad tokens at each step. This iterative process continues until every token position is determined.
As a fallback, if no token reaches the threshold $\tau$, we select the top-$\gamma$ tokens with the highest confidence score to guarantee progressive refinement of the denoising process.

\begin{table*}[t]

\caption{Comparison of different baselines on the test sets. LS denotes LibriSpeech, and CV denotes CommonVoice. The best and the second best results are shown in \textbf{bold} and by \underline{underlined}.}

\centering
\resizebox{\linewidth}{!}
{
\begin{tabular}{lccccccccccc} 
\toprule
\multirow{2}{*}{Model} & \multirow{2}{*}{Decoder Params} & 
\multicolumn{2}{c}{LS Clean} & \multicolumn{2}{c}{LS Other} & \multicolumn{2}{c}{CV Test} & \multicolumn{2}{c}{VoxPopuli} & \multicolumn{2}{c}{Average} \\ 
\cmidrule(lr){3-4} \cmidrule(lr){5-6} \cmidrule(lr){7-8} \cmidrule(lr){9-10} \cmidrule(lr){11-12} 
 & & WER(\%)$\downarrow$ & RTF$\downarrow$ & WER(\%)$\downarrow$ & RTF$\downarrow$ & WER(\%)$\downarrow$ & RTF$\downarrow$ & WER(\%)$\downarrow$ & RTF$\downarrow$ & WER(\%)$\downarrow$ & RTF$\downarrow$ \\
\midrule
Whisper-LLaMA3 & 8.03B & \textbf{2.15} & \underline{0.317} & 5.58 & 0.330 & \underline{8.55} & \underline{0.203} & 9.89 & \underline{0.268} & 6.54 & \underline{0.280 }\\
Whisper-Qwen3  & 8.19B & 2.72 & 0.410 & 6.62 & 0.427 & 9.18 & 0.355 & 10.06 & 0.364 & 7.15 & 0.389 \\
Whisper-LLaDA  & 8.02B & 2.34 & 1.678 & \underline{5.22} & 1.892 & 8.80 & 2.029 & \underline{9.68} & 1.344 & \underline{6.51} & 1.736 \\
dLLM-ASR       & 8.02B & \underline{2.28} & \textbf{0.057} & \textbf{5.17} & \textbf{0.076} & \textbf{8.36} & \textbf{0.057} & \textbf{9.56} & \textbf{0.060} & \textbf{6.34} & \textbf{0.063} \\
\bottomrule
\end{tabular}
}

\label{tab:asr_comparison}
\end{table*}

\section{Experiments}
\subsection{Datasets}
We train models on three widely used open-source datasets: LibriSpeech~\cite{librispeech}, CommonVoice-22.0-English~\cite{commonvoice}, and GigaSpeech~\cite{gigaspeech}, totaling 13,900 hours.
For evaluation, we assess the performance and efficiency of models on standard benchmarks, including the LibriSpeech test clean, LibriSpeech test other, and CommonVoice-22.0-English test set. 
Additionally, we incorporate the VoxPopuli~\cite{VoxPopuli} english test set as an out-of-domain benchmark to evaluate the model's generalization capability on unseen data distributions.

\subsection{Implementation Details}

We utilize the LLaDA-8B-Instruct~\footnote{\url{https://huggingface.co/GSAI-ML/LLaDA-8B-Instruct}} as our dLLM with 8 billion parameters. The text input is processed using the original LLaDA tokenizer.
We employ Whisper-large-v3's encoder~\cite{whisper} as the speech encoder, which remains frozen throughout the training process. The adapter comprises a 1D convolutional layer with a kernel size of 3 and a stride of 2, followed by two linear layers. The convolutional layer temporally downsamples the feature sequence, reducing the frame rate from 25 Hz to 12.5 Hz. Subsequently, the linear layer projects the speech features (d=1280) into the LLM's embedding space (d=4096).
During training, the total mask probability $\alpha$ is set as 0.2.
During inference, the confidence threshold is set to $\tau$, and the top-$\gamma$ selection parameter is set to 1.

All training and inference experiments are conducted on a cluster equipped with 16 NVIDIA A100 GPUs. 
The LoRA configuration is set to a rank of r=16, a scaling factor of 32, and a dropout rate of 0.05. 
We utilize the AdamW optimizer with $\beta_1$=0.9 and $\beta_2$=0.999 throughout the training process. 
In stage 1, we train the adapters with a total batch size of 256 for 5 epochs. We employ a cosine learning rate scheduler, where the learning rate increased linearly to a peak of $1 \times 10^{-4}$ over 4,000 warm-up steps before decaying. 
In stage 2, we jointly finetune the adapters and the LoRA-augmented backbone. We maintain the same hyperparameter configuration as the first stage, using a batch size of 256, 5 training epochs, and the identical learning rate schedule and warm-up strategy.

\begin{table}[t]
\centering
\caption{Results of ablation study on LibriSpeech clean and other testsets. LS denotes LibriSpeech.}
\label{tab:ablation}
\resizebox{\linewidth}{!}
{
\begin{tabular}{lcccc}
\toprule
\multirow{2}{*}{Model} & 
\multicolumn{2}{c}{LS Clean} & \multicolumn{2}{c}{LS Other} \\
\cmidrule(lr){2-3} \cmidrule(lr){4-5}
 & WER(\%)$\downarrow$ & RTF$\downarrow$ & WER(\%)$\downarrow$ & RTF$\downarrow$ \\
\midrule
dLLM-ASR                        & 2.28 & 0.057 & 5.17 & 0.076 \\
\quad w/o ASR Prior             & 2.29 & 0.069 & 5.20 & 0.084 \\
\quad w/o Length Pruning        & 2.27 & 0.071 & 5.13 & 0.089 \\
\quad w/o Chat-Style Prompt     & 2.87 & 0.056 & 5.76 & 0.076 \\
Whisper-LLaDA                    & 2.34 & 1.678 & 5.22 & 1.892 \\
\quad w Confidence-based Denoising        & 2.98 & 0.077  & 5.98 & 0.108 \\
\bottomrule
\end{tabular}
}
\end{table}

\subsection{Comparison Models}
To comprehensively evaluate the effectiveness of our proposed method, we compare our approach against representative models spanning distinct paradigms. First, we include AR baselines, specifically Whisper-Llama3 8B~\cite{llama3} and Whisper-Qwen3 8B~\cite{qwen3}, which couple the speech encoder with standard autoregressive language models.
Finally, we construct Whisper-LLaDA, which directly combines the speech encoder and an adapter with LLaDA, and serves as a baseline to quantify the specific contributions of our proposed strategies. Following~\cite{llada,speechdllm1}, Whisper-LLaDA uses a fixed generation length of 128 tokens, which covers 99\% of the speech transcripts.
All baselines are trained under exactly the same experimental setup, including the speech encoder, training steps, hyperparameters, and training data.

\subsection{Evaluation Metrics}

We evaluate model performance using two key metrics: recognition accuracy and computational efficiency. For accuracy, we adopt the word error rate (WER) as the evaluation standard on English test sets. For efficiency, we measure runtime using the real-time factor (RTF), defined as the ratio of compute time to input audio duration. This metric allows for a direct comparison of accuracy-efficiency trade-offs across different models. Average results are computed as the mean across test sets.

\subsection{Experiments Results}

As shown in Table \ref{tab:asr_comparison}, dLLM-ASR achieves the best balance between recognition accuracy and computational efficiency among all compared models. In terms of accuracy, dLLM-ASR demonstrates superior robustness, achieving the lowest WER on the more challenging LibriSpeech test other, CommonVoice test and VoxPopuli test sets, while maintaining competitive performance on LibriSpeech test clean. Regarding efficiency, despite having a parameter size ~8B comparable to Whisper-LLaMA3 and Whisper-Qwen3, dLLM-ASR significantly reduces the inference latency. It achieves an RTF of 0.063, representing a $4.44\times$ and $6.17\times$ speedup compared to Whisper-LLaMA3 and Whisper-Qwen3 baselines. 
The advantage is particularly prominent when compared directly with the Whisper-LLaDA baseline. 
Our model achieves a slight improvement in quality while drastically decreasing the RTF by 27.6 times.
This result highlights the ability of our model to deliver comparable ASR quality with $4.44\times$ speedup compared with Whisper-LLaMA3 baseline.

\begin{figure}[t!]
  \centering
  \includegraphics[width=1\linewidth]{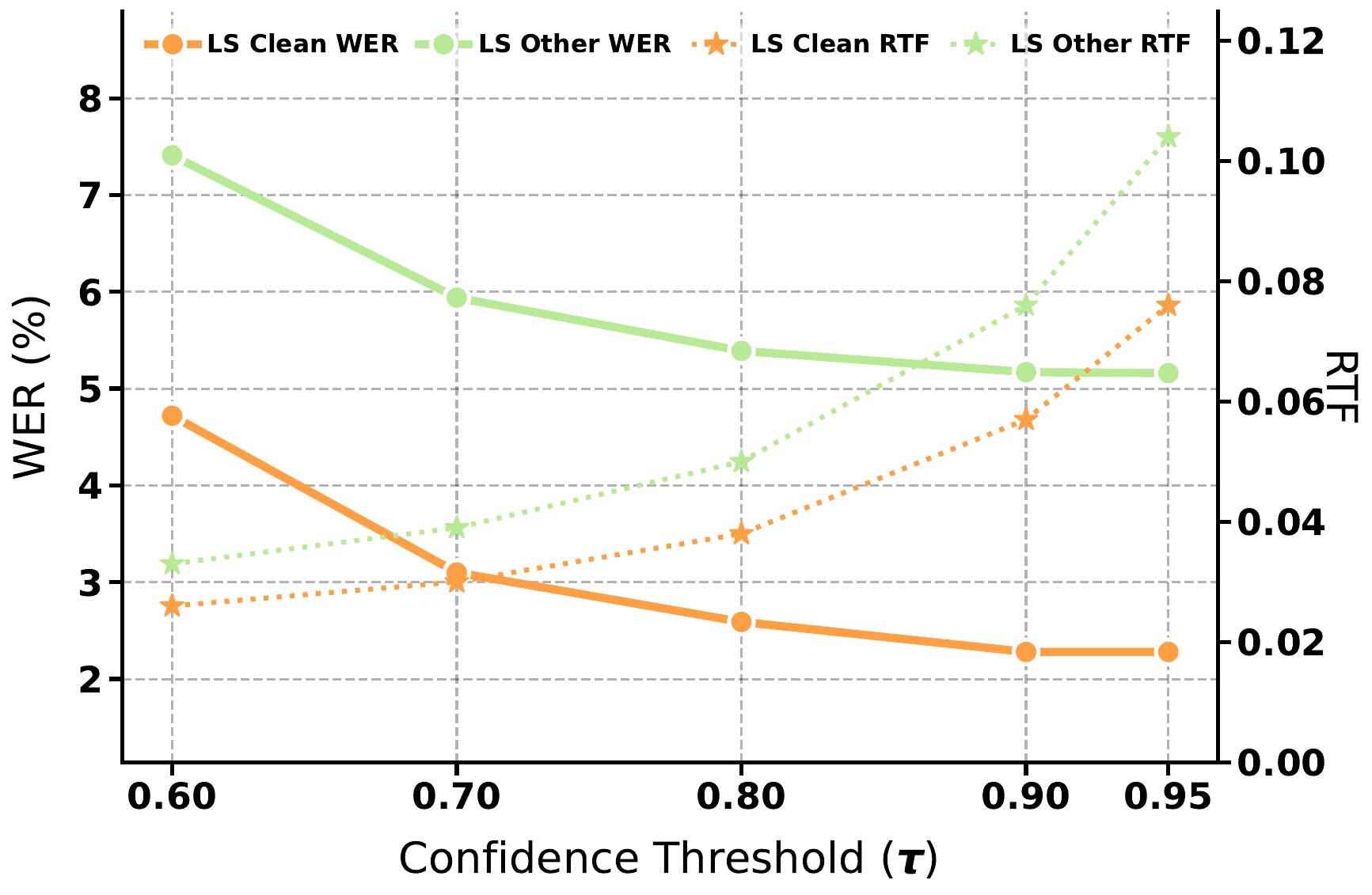}  
  \caption{Ablation study on the confidence threshold ($\tau$) for the confidence-based allocation strategy. The results are reported on LibriSpeech (LS) clean and other test sets.}
  \label{fig_confidence}
\end{figure}

\subsection{Ablation Studies}
To validate the contribution of each component, we conduct an ablation study on the LibriSpeech clean and other test sets, as shown in Table~\ref{tab:ablation}. 
Removing the ASR prior results in an increase in RTF, demonstrating that it's essential for guiding the diffusion process and avoiding the all-mask start.
In our experiments, we observe that this prior significantly increases the number of tokens whose confidence exceeds the predefined threshold $\tau$ at the first denoising step, thereby accelerating the overall denoising process.
Our pruning strategy also successfully removes redundant tokens to boost speed.
Moreover, ablating the chat-style prompt described in Section~\ref{training_strategy} leads to a noticeable drop in generation quality, suggesting that it may facilitate better cross-modal alignment.
To investigate the effect of the confidence-based denoising strategy alone, we apply it directly to the Whisper-LLaDA baseline. It successfully reduces latency (RTF = 0.077) but causes a decline in accuracy. 
This result demonstrates that inference acceleration cannot be achieved in isolation. It requires the support of our proposed strategies to maintain robustness at high speeds.
In summary, the full dLLM-ASR framework achieves the best trade-off between recognition accuracy and inference speed, validating the synergy of proposed components.


Furthermore, to determine the optimal settings for the confidence-based denoising strategy, we conduct an experiment on the confidence threshold $\tau$.
As summarized in Fig.~\ref{fig_confidence}, we evaluate dLLM-ASR under various threshold configurations. 
The results reveal a clear trade-off between efficiency and performance. As the confidence threshold decreases from 0.6 to 0.9, the token acceptance rate increases, allowing the model to determine tokens earlier. Obviously, this leads to lower RTF but higher WER. Conversely, a higher threshold ensures quality but diminishes the speedup gains. 
We observe that the WER corresponding to $\tau= 0.9$ and $\tau= 0.95$ exhibits negligible differences, making $\tau= 0.9$ the optimal choice for balancing inference speed and recognition precision.
Consequently, we adopt $\tau=0.9$ as the default setting for our final proposed framework.

\section{Conclusion}
\label{sec:conclusion}

In this paper, we introduce dLLM-ASR, a novel framework designed to bridge the gap between the high accuracy of AR models and the inference efficiency of NAR approaches.
We mitigate the mismatch when applying dLLM into speech recognition, reformulating the discrete diffusion decoding as a prior-guided and adaptive refinement process.
Specifically, we introduce a speech's ASR prior to guide the denoising process and anchor generation length. And we use length-adaptive pruning to dynamically adjust sequence length and confidence-based early exit to allocate computation only where needed.
With the prior-guided and adaptive denoising, we have reduced a large amount of computational redundancy.
Experimental results demonstrate that dLLM-ASR establishes a new Pareto frontier in speech recognition, delivering accuracy comparable to LLM-ASR models while achieving a 4.44$\times$ inference speedup. Future work will focus on extending this diffusion-based paradigm to streaming ASR scenarios and tasks.

\bibliographystyle{IEEEbib}
\bibliography{icme2026references}


\end{document}